\documentclass[conference]{IEEEtran}
\IEEEoverridecommandlockouts
\usepackage{cite}
\usepackage{amsmath,amssymb,amsfonts}
\usepackage{algorithmic}
\usepackage{graphicx}
\usepackage{textcomp}
\usepackage{xcolor}
\def\BibTeX{{\rm B\kern-.05em{\sc i\kern-.025em b}\kern-.08em
    T\kern-.1667em\lower.7ex\hbox{E}\kern-.125emX}}
\begin{document}

\title{Empowering IoT Applications with Flexible, Energy-Efficient Remote Management of Low-Power Edge Devices }

\author{\IEEEauthorblockN{1\textsuperscript{st} Shadi Attarha}
\IEEEauthorblockA{\textit{Dept. Communication Networks} \\
\textit{University of Bremen, Germany}\\
sattarha@uni-bremen.de}
\and
\IEEEauthorblockN{2\textsuperscript{nd} Anna Förster}
\IEEEauthorblockA{\textit{Dept. Communication Networks} \\
\textit{University of Bremen, Germany}\\
anna.foerster@uni-bremen.de}

}

\maketitle

\begin{abstract}
In the context of the Internet of Things (IoT), reliable and energy-efficient provision of IoT applications has become critical. Equipping IoT systems with tools that enable a flexible, well-performing, and automated way of monitoring and managing IoT edge devices is an essential prerequisite. In current IoT systems, low-power edge appliances have been utilized in a way that can not be controlled and re-configured in a timely manner. Hence, conducting a trade-off solution between manageability, performance and design requirements are demanded. 
This paper introduces a novel approach for fine-grained monitoring and managing individual micro-services within low-power edge devices, which improves system reliability and energy efficiency. The proposed method enables operational flexibility for IoT edge devices by leveraging a modularization technique. Following a review of existing solutions for remote-managed IoT services, a detailed description of the suggested approach is presented. Also, to explore the essential design principles that must be considered in this approach, the suggested architecture is elaborated in detail. Finally, the advantages of the proposed solution to deal with disruptions are demonstrated in the proof of concept-based experiments.
\end{abstract}


\section{Introduction}
 \label{sec:intro}
In recent years, Internet-of-Things (IoT) has attained increased interest and has been considered a promising method in various fields such as industry, agriculture, and healthcare. The main aim of leveraging IoT is to collect data from distributed edge devices, process it, and provide timely information to end-users for decision-making\cite{ieee2015towards}. This process helps in system performance enhancement and cost-effectiveness. In common IoT systems, to decrease the network overhead, different tasks are performed on the edge devices, all of which are designed with extremely limited resources \cite{zandberg2022femto}. Therefore, ensuring data reliability, resource efficiency, and system transparency is of utmost importance.

To ensure the reliable and efficient delivery of IoT services within low-power edge devices, it is crucial to prioritize operational flexibility and manageability.
However, common IoT software deployments are unstructured and do not offer fine-grained monitoring and managing support over individual services remotely. For example, one typical problem is when one sensor becomes faulty for some reason, and the node either continues to deliver incorrect data and waste energy or all other services also crash.
Therefore, the possibility of managing failure or misbehavior to protect IoT systems against them is very challenging \cite{attarha2023automated}. Additionally, state-of-the-art solutions for remote monitoring are resource-demanding and not suitable for edge devices with limited resources.
This attitude creates bottlenecks that impact flexibility and reliability in IoT applications. 

In this context, service isolation is a strategy that provides a level of modularity required for observing and controlling individual services. 
This paradigm improves operational flexibility and energy utilization by enabling agile and more straightforward maintenance procedures.
In other words, in the event of malfunctions, the related part (e.g., a faulty sensor) can be reconfigured and managed from a remote location in a timely manner.

Recent research focuses on using containerization technology such as LXC and Docker, along with Single Board Computers as IoT gateways, to achieve on-demand control of services in IoT systems. However, due to the limited resources of IoT edge devices, these solutions are not feasible there. In this regard, the lack of a systematic method for monitoring and controlling IoT services deployed in edge devices equipped with extremely constrained resources leads to a challenge and a requirement for further investigation in this area. In particular, it is vital to provide fast and efficient controlling methods that enable monitoring and securely managing micro-services within low-power IoT devices. 

This paper presents a novel approach to improve service maintenance procedures, resource utilization, and system reliability for IoT edge devices characterized by extremely limited resources. 
The proposed approach empowers IoT edge devices with the possibility to run multiple isolated services, where there is an ability to have control over an individual service. This capability is precious as it allows for independently managing a faulty component (e.g., a broken sensor ) within an edge device without compromising the other components' functionality (i.e., remaining sensors can continue working without disruption). The resulting operational flexibility will be achieved by utilizing a modularization technique. A literature survey on the manageability of IoT edge devices is discussed in Section \ref{Sot}. Then, the proposed solution, including its envisioned architecture and operational states, is described in detail. Finally, an experiment-based proof of concept is presented to demonstrate the impact of the proposed approach on an IoT-based system. 

\section{State-of-the-art and Research Challenge}\label{Sot}

Two vital steps come into play to facilitate the independent management of IoT services and minimize the effects of disruptions: real-time monitoring and remote management capability. Given the current work focuses on independent manageability, this section explores state-of-the-art pertaining to this topic.

To achieve the manageability of individual services while maintaining the functionality of other services, modularization is a critical step. Several noteworthy research has been pursued to enable the isolation of services deployed on low-power devices. 
\textit{Darjeeling} \cite{brouwers2009darjeeling} and \textit{Velox} \cite{tsiftes2018velox} are compact Virtual Machines (VM) that isolate services on MCUs. However, their deployment in IoT edge devices is challenging due to the significant increase in resource consumption through implementing their required libraries. 

In a recent study \cite{zandberg2022femto}, authors introduced the \textit{Femto} container, an extremely compact VM integrated with the RIOT operating system. The primary aim of this approach is to provide process isolation and enable remote updating. The authors have presented that multiple services can be effectively deployed as containers. Additionally, they have highlighted the availability of remote device updates, although it is important to note that managing services foresees updating the underlying software. This might cause a disruption in other services' functionalities by recompiling the underlying software and other services.  
Toit \footnote{https://toit.io/} is a platform and language designed to simplify IoT application development and management. It includes a cloud-based platform for managing services individually. Using Toit enables securing the code on MCU with lightweight containers where each container can be managed independently. In this work, we will use Toit as a tool for implementing our proposal.

Investigating the previous research in low-power IoT edge devices indicates that the current literature mainly focuses on isolating services (i.e., assuring each service can not access memory regions outside what is allowed). However, the possibility of managing IoT services individually without compromising other services' functionality, essential for system reliability and cost-effectiveness, still needs to be studied. In this paper, we explore the approach for fine-grained monitoring and managing individual IoT services, which provides the ability to decide and react to disruptions adaptively.

\section{Proposed Solution }
We assume an IoT system consisting of many edge devices, where each device hosts several micro-services, such as sensor readings, sensor data analysis, etc. 
Our goal is to enable real-time monitoring of each micro-services and a remote management function to enable/disable them as well as repair/update faulty services, where there is no requirement for recompiling the underlying software and other services.
To ensure a reliable and energy-conserving  IoT system that utilizes low-power edge devices, it is necessary to focus on the following two considerations:
\begin{itemize}
    \item Observability: Ensuring that the edge device can be monitored and its performance can be evaluated is essential for maintaining IoT systems' reliability and energy efficiency. This includes the monitoring of an edge device and detecting issues that indicate that either a sensor, an MCU, or software is not working correctly.    
    \item Manageability: Efficient management of edge devices is essential to ensure optimal IoT system performance. This includes remotely configuring and updating in response to real-time concerns.
\end{itemize}

This work focuses on the manageability issue and aims to develop a flexible and modular framework for IoT applications, allowing independent management of micro-services.

\begin{figure}[t]
	\centering
	\centerline{\includegraphics[scale=0.88]{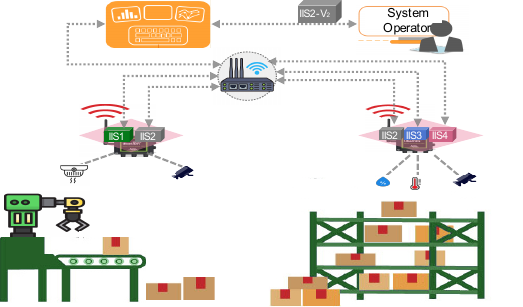}}
	\caption{Proposed solution applied in a smart warehouse}
	\label{fig_general}
\end{figure}

Figure \ref{fig_general} illustrates the application of our approach in an IoT system, considering a smart warehouse scenario. In this approach, IoT services are deployed on isolated entities such as containers, micro-services, or virtual machines, collectively called Isolated IoT Services (IIS). For instance, IIS3 gathers temperature sensor data and is manageable independently. The subsequent sections will explain the proposed architecture, highlighting its structural aspects and defining the operational states of the IIS to assess its performance.

\subsection{Proposed Architecture}
To promote essential design components for deploying and managing IISs over the available low-power edge devices, Figure \ref{fig_arch} presents the proposed architecture. It consists of three abstract layers: \textit{Infrastructure}, \textit{Service}, and \textit{Control}. The \textit{Infrastructure} layer hosts the software (i.e., firmware) and hardware (i.e., Processor, memory, I/O) resources that enable the operational environment for running IoT services. The \textit{Service} layer hosts the IIS, which is a software entity that represents the implementation of a current isolated IoT service. IISs can be deployed inside lightweight virtual machines or containers designed for low-power edge devices (e.g., Toit). The \textit{Control} layer is the key part of the architecture and has the responsibility for observing and managing IISs and edge devices to fulfill system requirements and performance goals. Three different modules, namely: \textit{Data monitoring},  \textit{Anomaly detection}, and \textit{Service orchestration}, collaborate to collect the measurements, detect abnormal behaviors and provide reasonable reactions for maintaining IoT system performance. Continuously gathering sensor measurements and preparing them (e.g., formatting, cleaning, and sampling) for the \textit{Anomaly detection} module are covered by the \textit{Data monitoring} module.  
The \textit{Anomaly detection} module detects anomalies in the gathered data, such as an outlier or drift, and makes notifications to \textit{Service orchestration}. Statistical methods and machine learning-based anomaly detection models can be utilized to recognize abnormal behaviors \cite{sgueglia2022systematic}. Here, once could consider either a fully edge-based solution or an edge/fog/cloud distributed one to optimize resources. \textit{Service management} provides actions related to individual IoT services' life-cycle, such as starting, stopping, and reprogramming.

\subsection{Operational State of Isolated IoT Service}
IoT services can be prone to a number of faults that can result in either complete or partial failures. In order to analyze the performance of the IIS, two main metrics are considered – \textit{Availability} and \textit{Correctness} \cite{sterbenz2010resilience}. These metrics assess disruptions' impact on IIS and evaluate system performance. 
\textit{Availability} evaluates the presence of sensor measurements at a given time, which is the input for the anomaly detection module. \textit{Correctness}, assessed by the \textit{Anomaly detection} module through its data analysis tasks, evaluates the state of collected measurements as either faulty or correct. It reflects the trustworthiness of the data, which directly affects the accuracy of the results produced by the IoT system. 
Based on the aforementioned criteria, the operational state of IIS can be determined as follows: 
\begin{figure}[t]
	\centering
	\centerline{\includegraphics[scale=0.83]{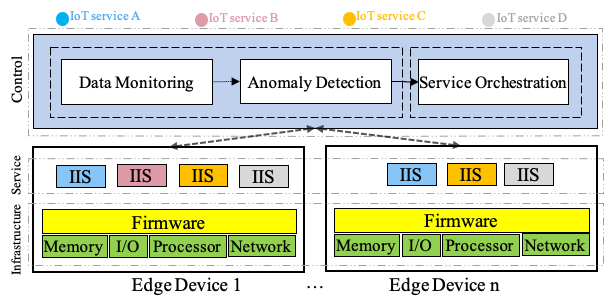}}
	\caption{Proposed architecture }
	\label{fig_arch}
\end{figure}

\textit{Normal state}: For an IoT service, IIS\textsubscript{x}, when all above metrics are met for its collected measurements, the IIS\textsubscript{x} is considered fully functional, and the measurements are accurate. Hence, available measurements are analyzed to gain awareness about its state.

\textit{Suspicious state}: Disruptions in the IoT system may cause the IoT service (i.e., IIS) to be no longer practical, i.e., \textit{Availability} and/or \textit{Correctness} are not satisfied. This means an IIS either collects inaccurate measurements or is unavailable, which would be determined by \textit{Data monitoring} and \textit{Anomaly detection} modules. As a consequence, it might lead to wrong decisions. This way, the system operator can get a real-time understanding of the performance of the system. 

Regarding the state of IIS, the performance of the corresponding sensor/task can be observed, and thus disruptions can be handled timely by taking appropriate control actions over a faulty component. For example, when the \textit{Availability} metric is not satisfied and causes the \textit{Suspicious state} for an IIS\textsubscript{x}, the system administrator can first stop it without modifying other IISs or recompiling underlying software, causing reduced energy consumption and resource usage. Then, by finding the root cause, the proper control action to tun the IoT system can be taken. Meanwhile, it's important to note that all other IISs continue to function flawlessly and experience no interruptions during this period.

\section{Proof of concept}
This section contains an experiment-based proof of concept to explore the advantages of the proposed solution in case of an anomaly in an exemplary IoT system. In this study, energy consumption is considered as a performance metric that is vital for resource-constraint (i.e., battery-based) devices.

In these experiments, we try to emulate an IoT-based system, where the IoT edge device is composed of one temperature-humidity sensor, DHT11, and a carbon dioxide sensor, SCD30, all connected to ESP-based WROOM-32's I/O pins. 
To isolate IoT services in our system, we utilized the Toit platform as discussed in Section \ref{Sot}. By isolating IoT services, we separated each service from one another and the system firmware, which is critical for ensuring the security and dependability of the services. In this scenario, two different services collect data from specific sensors. One service is dedicated to gathering data from the DHT11 sensor, while the other is focused on collecting data from the SCD30 sensor. In the proposed approach, upon detecting a deviation, we have the ability to stop the corresponding service while keeping the system working without any disruption to avoid wasting resources and producing inaccurate values, resulting in system unreliability. We emulate problems with one of the sensors to demo the proposal. 

\subsection{Energy consumption evaluation}
Evaluating energy consumption in the IoT domain is challenging due to hardware dependencies and various parameters\cite{ruckebusch2018modelling}. For our experiments, we measure the energy consumption of IoT edge devices using a USB multi-meter tester. 
To validate measurement accuracy, the obtained values were cross-checked with the hardware's datasheet for plausibility, detecting any errors or discrepancies. 

To demonstrate the impact of the proposed approach on energy consumption, we conducted two tests where the device wakes up every 10 minutes to measure and transmit data using the WiFi interface to an internet service using the MQTT protocol. The experiment was carried out over six hours. In the scenario for both tests, during the first three hours, there is an assumption that sensors and the MCU (i.e., ESP32) were functioning correctly and accurate data values were obtained. Then, it was assumed to have a disruptive event (e.g., covering by external objects or a broken sensor) for the CO\textsubscript{2} sensor caused the values to be inaccurate for the following hours. 
In the first test, upon happening the deviation, the \textit{Suspicious} state for a service related to SCD30 arises. Hence, it promotes mitigation actions to be taken to prevent system unreliability and resource wastage, given the energy-intensive nature of the corresponding sensor (i.e., SCD30). Therefore, the service related to the CO\textsubscript{2} sensor has been stopped by the \textit{Service Orchestrator} entity of our architecture in the first test and is uninstalled from ESP32. This procedure is executed seamlessly, ensuring the uninterrupted operation of another service without any disturbances. Accordingly, just a service related to DHT11 was operating in MCU for the next hours. In the second test, the system's energy usage was evaluated in a situation where fine-grained control over individual services could not be possible. In this case, the code was executed natively on the ESP32, representing the traditional IoT system development approach. Since there is no possibility of managing the services, although the carbon dioxide sensor has a problem, the device continues sending the (faulty) data. This results in wasting energy and inaccurate measurements.

\begin{figure}
  \centering
  \includegraphics[width=0.5\textwidth]{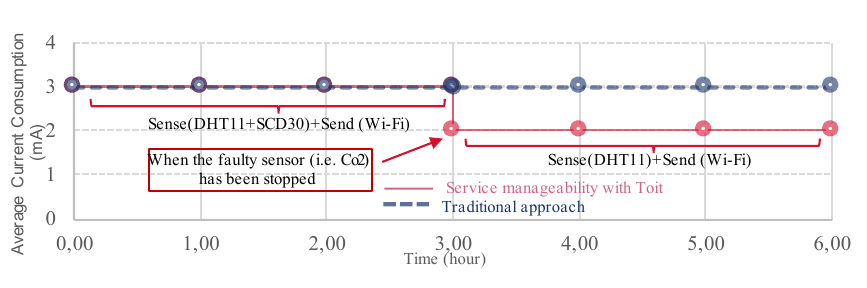}
  \caption{Comparison of the average current consumption of an IoT device considering service manageability and traditional method. Applying the proposed approach by leveraging Toit, the power consumption is reduced by, on average, 1 mA leading to a longer lifetime while keeping the system functional and reliable.}
  \label{fig_energy_good}
\end{figure}

The experiment's results are displayed in Figure \ref{fig_energy_good}, showcasing the impact of manageability on energy consumption within low-power IoT devices. By comparing the average current between the two tests, valuable insights are gained regarding the effectiveness of managing individual services and optimizing energy usage. Furthermore, it enables the system to continue operating even if it is not fully functional while still maintaining a level of reliability.
The experiment outcomes reveal that Toit's overhead in this scenario is negligible, positioning it as a viable solution for service isolation. This finding is particularly significant for IoT devices characterized by limited resources and reliance on battery power.

\subsection{Discussion}

The results of our study highlight the significance of manageability for low-power edge devices within an efficient IoT system. Our proposed approach offers the advantage of independent service control without necessitating changes to the underlying device software, such as firmware. This eliminates the need for substantial data transfer to update a new firmware and mitigates the adverse impact on energy consumption caused by the size of updates. Consequently, our method presents a viable alternative to the firmware-over-the-air update technique, making it a preferable choice in this context.
However, it is important to carefully manage the overhead associated with modularization techniques (e.g., Toit) to prevent excessive energy consumption. Finding the right balance between specific performance targets and service manageability is crucial to optimize the advantages of the proposed approach while minimizing.

\section{Conclusion }
This paper has introduced a novel approach that empowers low-power IoT devices with increased operational flexibility and observability. This results in independent service management at run-time without affecting the overall operation of other services.
The effectiveness of this approach in mitigating disruptions has been evaluated through a real-life experiment. As the future research direction, the proposed architecture and its building blocks will be investigated regarding potential deployment approaches and interfaces. Furthermore, potential countermeasures to address faults through remote management will be analyzed. Studying optimizing modularization techniques in low-power IoT devices should also be essential.

\bibliographystyle{IEEEtran}

\bibliography{Reference.bib}

\end{document}